# Mathematical Modeling of Plasticity and Heterogeneity in EMT

Shubham Tripathi[1], Jianhua Xing[2, *], Herbert Levine[3, 4, *], and Mohit Kumar Jolly[5, *]

[1] PhD Program in Systems, Synthetic, and Physical Biology, Rice University, Houston, TX 77005, USA.

[2] Department of Computational and Systems Biology, University of Pittsburgh, Pittsburgh, PA 15261, USA.

[3] Department of Physics and Department of Bioengineering, Northeastern University, Boston, MA 02115, USA.

[4] Center for Theoretical Biological Physics, Rice University, Houston, TX 77005, USA.

[5] Centre for BioSystems Science and Engineering, Indian Institute of Science, Bangalore, Karnataka 560012, India.

* Email: xing1@pitt.edu (J.X.), h.levine@northeastern.edu (H.L.), mkjolly@iisc.ac.in (M.K.J.).

## Abstract

Epithelial-mesenchymal transition (EMT) and the corresponding reverse process, mesenchymal-epithelial transition (MET), are dynamic and reversible cellular programs orchestrated by many changes at biochemical and morphological levels. A recent surge in identifying the molecular mechanisms underlying EMT / MET has led to the development of various mathematical models that have contributed to our improved understanding of dynamics at single-cell and population levels: a) multistability (how many phenotypes can cells attain en route EMT / MET?), b) reversibility / irreversibility (what time and / or concentration of an EMT inducer marks the 'tipping point' when cells induced to undergo EMT cannot revert?), c) symmetry in EMT / MET (do cells take the same path when reverting as they took during the induction of EMT?), d) non-cell autonomous mechanisms (how does a cell undergoing EMT alter the tendency of its neighbors to undergo EMT)? These dynamical traits may facilitate a heterogenous response within a cell population undergoing EMT / MET. Here, we present a few examples of designing different mathematical models that can contribute to decoding EMT / MET dynamics.

## Introduction

Epithelial-mesenchymal transition (EMT) is a cellular process involving changes in multiple aspects of cellular behavior— cell-cell adhesion, cell polarity, cell migration and invasion, and cell shape[1]. EMT and the corresponding reverse process, mesenchymal-epithelial transition (MET), are regulated at multiple levels. These include transcriptional, post-transcriptional, translational, and epigenetic[2] controls, along with non-cell autonomous mechanisms acting through matrix density[3] or cell-cell communication[4–7]. Largely thought of in the past as a binary process, EMT is now known to involve multiple stable intermediates referred to as hybrid epithelial / mesenchymal (hybrid E / M) phenotypes[8]. This updated view of the process has, in part, been driven by predictions made by various mathematical models for the regulatory networks involved in EMT[4,5,9–14]. These mathematical models have focused on characterizing the properties of EMT and have predicted that cells can stably maintain one or more hybrid E / M phenotypes[15]. Moreover, these models have also driven insights into how cells may spontaneously switch among various phenotypes due to stochasticity, and thereby determine how cellular plasticity leads to phenotypic heterogeneity associated with EMT as observed experimentally[16,17]. These models have also offered mechanistic insights into experimental observations showing that EMT and MET are not necessarily symmetric processes[12,18], i.e., cells may take different paths during EMT and MET in the multi-dimensional landscape of epithelial-mesenchymal plasticity. Finally, these models have helped us gain insights into the interconnection between EMT and other cellular traits such as stemness; for instance, the prediction that a hybrid E / M phenotype is more stem-like and metastatically aggressive than cells exhibiting extremely epithelial or extremely mesenchymal phenotypes[19] was recently confirmed both *in vitro* and *in vivo*[20–22]. Here, we introduce a generic framework for developing mathematical models of EMT regulation and share examples of how these models can be used as tools to generate predictions that will guide the next set of experiments.

# Mathematical Modeling of EMT

The choice of a systems biology approach to study a biological process is highly context-dependent. We here describe a generic procedure for choosing an appropriate approach and detail how this procedure was applied to modeling epithelial-mesenchymal transition (EMT).

**Identify a problem that mathematical modeling can help address and form a team of experimental and modeling researchers.**

This is a key and probably the most challenging step in modeling studies. There are questions that modeling studies can address and others that they cannot address. It is typically constructive to form a team of experimental and modeling researchers. The team members hold thorough literature review and extensive, in-depth discussions to review existing knowledge and identify open questions regarding the system. One may find it pedagogically illuminating to read accounts of how some successful collaborations were established[23,24].

**Choose an appropriate modeling framework.**

Several modeling frameworks have been used to analyze EMT regulatory networks. A Boolean network has dynamics that are discrete in time and involve discrete variable values. The variable values are updated based on a set of Boolean functions that reflect the regulatory relations[25]. Conversely, an ordinary differential equation (ODE)-based model treats time and variables as taking continuous values. Both Boolean network and ODE-based models can be deterministic (meaning that one can precisely predict the temporal evolution of the variables from a set of initial conditions), or stochastic (meaning that the prediction is only probabilistic). There is no best modeling framework for all cases, and one needs to determine what is appropriate and justified for the biological system and process under study. Some general aspects that may be considered include:

1. What is the qualitative and quantitative information available? Compared to a Boolean model, an ODE model typically has more parameters and requires more quantitative data to constrain these model parameters. Therefore, for a large regulatory network without much quantitative data such as the one studied by Steinway *et al.*[11], the Boolean framework is appropriate. It would be questionable whether an alternative ODE-based model with dozens or even hundreds of free parameters can provide further additional information (but see systematic statistical analyses of model ensembles discussed below).
2. Is the framework sufficient to describe the system dynamics, and does it provide new mechanistic insights that would be unavailable or unclear without the modeling approach? Each framework has its limitations. For example, a Boolean model typically uses some universal parameters and only provides qualitative or at most semi-quantitative information. It can be a good starting point to analyze how multiple regulatory factors interact to generate different EMT cell types as demonstrated by Steinway *et al.*[11]. The model has limited capacity to describe how different time scales of the signal transduction pathways involved in EMT contribute to quantitative detection and encoding of the dose and duration information of the stimulating signals. For the latter purpose, an ODE-based model is a more appropriate choice, as demonstrated by Zhang *et al.*[26] to show how pathway crosstalk leads to a temporal check-point mechanism for detecting TGF-$\beta$ duration information.

As a rule of thumb, one chooses a modeling framework that it is simple and sufficient to address the underlying problem. The widely regarded criterion suggested by Einstein for evaluating physics theories also applies here: "Everything should be made as simple as possible, but not simpler". It is possible that for a given problem, initially a coarse-grained framework is appropriate, and as more and more quantitative data becomes available, a different framework becomes necessary to incorporate the new information.

Unfortunately, a commonly held misconception emphasizes that it is always desirable to incorporate additional biological details explicitly into a mathematical model, and this tendency is further reinforced by the expanding computational power. However, abstraction is necessary and is done in all modeling efforts. We want to stress here that the most important reason for using modeling approaches is to provide mechanistic insight buried in the data, and not just to crank machines and obtain some numbers. For this purpose, it is both productive and necessary to perform proper abstraction and idealization as successfully used in theoretical physics[27]. A simple model that only makes qualitative predictions but provides deep mechanistic insight has more value than a complex model that can only "reproduce" experimental data but does not necessarily make a new set predictions that may be tested experimentally to improve our understanding of the system. To be fair, both detailed and simplified approaches have their merits, and sometimes it is constructive to combine the two strategies. One may start with detailed models that can reproduce the data, then remove model ingredients step-by-step to identify the minimal components that are essential for recapitulating the key dynamical features of the system.

**Construct a mathematical model and perform analysis.**

With the problem identified and an appropriate modeling framework selected, one can follow some generic modeling procedures:

1. Summarize known interacting species into a regulatory network. If there are uncertain interactions, one may construct a set of possible networks for later comparative studies. Fig. 1 shows a core EMT regulatory network used in several studies[9,13,28].
2. Set up mathematical equations based on the biology. This step is nothing more than translating the relevant biological information into mathematical forms. For example, the equation below governs the temporal evolution of the total level of SNAIL1 mRNA ($[snail1]_t$), which is summed over both free ($[snail1]$) and miR-34 bound ($[snail1]_t - [snail1]$) mRNAs[28].

$$\frac{d[snail1]_t}{dt} = \underbrace{k_0}_{\text{basal expression}} + \underbrace{k \frac{([TGF]_t / K_1)^2}{1 + ([TGF]_t / K_1)^2}}_{\text{TGF-}\beta \text{ activation}} \underbrace{\frac{1}{1 + [SNAIL1]/K_2}}_{\text{SNAIL1 self-inhibition}} - \underbrace{k_{d0}[snail1]}_{snail1 \text{ basal degradation}} - \underbrace{k_d ([snail1]_t - [snail1])}_{miR-34 \text{ regulated } snail1 \text{ degradation}}$$

   Each term on the right-hand side of the above equation corresponds to one of the SNAIL1 related links in fig. 1.
3. Constrain model parameters using the available quantitative data. Several parameter estimation algorithms are available, from linear regression to the more sophisticated maximum likelihood estimation, and Markov chain Monte Carlo methods. Since, in practice, it is rare to have sufficient data for a specific system under study, a commonly adopted practice is to estimate the many parameters based on data from different labs, different cell lines, or cells from different tissues. However, even results from the same cell line can be quantitatively different due to factors such as difference in cell generation, reagent vendors, or even batches. Besides, dynamic parameters such as mRNA turnover rates can differ by orders of magnitude for cells under different conditions. An emerging trend is to collect data from one lab or under the same experimental settings[29], similar to what has been adopted in some large consortiums like ENCODE. Furthermore, instead of using only the best-fit parameter set, one may use an ensemble of model parameters to make model predictions. Zhang *et al.*[26] adopted such an integrated modeling-quantitative measurement procedure and an ensemble-based approach has been developed previously[30,31]. Another model ensemble method is discussed in the next section.
4. Specify initial conditions (e.g., initial concentrations of various species) that reflect the experimental setup. For example, if one models cell response after adding TGF-$\beta$ at time *0*, one may first make a rough estimation of the initial concentrations, then propagate the ODEs for a sufficiently long time to reach a steady state, and use the steady state values as the initial conditions at time *0*.
5. Perform standard analyses such as bifurcation analysis, phase diagram, temporal trajectories, and robustness / sensitivity analysis. One may either write custom computer code (e.g., in Matlab, Python, etc.), or use available computer packages, e.g., XPP (http://www.math.pitt.edu/~bard/xpp/xpp.html), Oscill8 (http://oscill8.sourceforge.net/), and BioNetGen[32]

**Explain available experiments and make testable predictions.**

A unique advantage of computational modeling over experimental studies is that, generally, it is much easier to perform a series of *in silico* studies than their experimental counterparts as the latter may be either time and resource consuming, or may even not be feasible. Generally speaking, mathematical / computational modeling can

1. Provide mechanistic insights not evident from the data, and sometimes resolve conflicting experimental results or distinguish competing mechanisms. For a given system, data are typically collected from different sources and using different techniques. Each experimental technique or approach can only reveal partial information about the system, and modeling integrates the discrete information. By placing all the experimental results on a common ground, a modeling study allows one to check whether the data are consistent mutually, and with the conceived mechanisms.

2. Make predictions leading to new experimental measurements that might not have been considered otherwise. For example, the modeling study by Tian et al.[13] inspired a subsequent measurement of single cell SNAIL1 expression levels using flow cytometry[28].

3. Identify essential ingredients or missing links necessary to explain the observations. For a given system, there may be too much information, and some of it may not be or may only be marginally relevant to addressing a specific question. By adding or removing individual components and examining the effect on model behavior, one can identify the essential ingredients of a model. In other cases, the available information may be insufficient. In such a scenario, following a similar procedure of systemically adding individual components, one can predict the missing component(s) that are necessary to explain the experimental results. The missing component may then be identified in subsequent experimental studies. For example, the study by Lu *et al.*[9] suggested the existence of positive feedback in the regulation of ZEB in EMT regulation (dashed line in fig. 1).

It is important to point out that a model need not necessarily be right in order to be useful. In fact, every model is only an approximation and abstraction of the biological system under study and will be replaced by better approximations when additional information becomes available. "All models are wrong, but some are useful"[33]. Even an eventually falsified model may suggest useful experimental studies that would otherwise not have been performed, and thus help in advancing our knowledge of a biological system. Such models should receive deserved credit.

**Perform corresponding experimental studies.**

As Katchalsky pointed out[24], "Theory tells us what cannot happen, and it can tell us what could happen. But only experiments tell us what does happen." All model predictions need to be subject to subsequent experimental tests.

**Go back to step 2 and iterate; expansion of model (even after publishing the original work).**

It has become more and more common to see studies that have iterations between modeling and experiments. Sometimes the integrated experiment-modeling process may even lead to revisiting step 1 to define new questions and seek expanded collaborations. For example, early modeling studies[9,13] on EMT focused on the core regulatory network (fig. 1). Several subsequent studies expanded the network to explore how additional factors contribute to the spectrum of EMT phenotypes[4,34,35].

# Modeling Population Heterogeneity in EMT

Intra-tumoral heterogeneity, wherein cancer cells within the same tumor exhibit different phenotypes, has been reported across multiple cancer types, both *in vitro* and *in vivo*[36]. Tumor cell populations in different cancer types including leukemia[37], breast cancer[38], colorectal cancer[39,40], brain cancer[41], and prostate cancer[42] can consist of subpopulations of cells that exhibit stem cell-like behavior. Cells in triple-negative breast cancer can exhibit distinct phenotypes including luminal, basal, immunomodulatory, mesenchymal, and stem-like[43]. In small cell lung cancer, tumor cells can exhibit both neuroendocrine and non-neuroendocrine phenotypes[44]. Intra-tumoral heterogeneity has recently been identified as a principal cause for the failure of anti-cancer therapies[45]. Therefore, characterization of the mechanisms driving this feature of tumor cell populations is key to advancing anti-cancer therapeutics. In many (perhaps most) cases, genetic heterogeneity does not underlie phenotypic heterogeneity, i.e., tumor cells exhibit different phenotypes in spite of carrying the same genetic alterations. This indicates that non-genetic mechanisms may be the chief driver of intra-tumoral heterogeneity.

Cells within the same tumor can exhibit different EMT-associated phenotypes— an epithelial phenotype, a mesenchymal phenotype, and one or more hybrid E / M phenotypes. This is a canonical example of non-genetic intra-tumoral heterogeneity observed across cancer types including in breast cancer[46], melanoma[47], colorectal cancer[48], and in prostate cancer[49]. Different EMT-associated phenotypes exhibit varying tumor-initiating capabilities[6,7] and sensitivities to anti-cancer drugs[50,51]. How does such epithelial-mesenchymal heterogeneity emerge in a population of cancer cells? How is this heterogeneity maintained and propagated across generations and passages? These are key questions that must be answered if we are to be able to attenuate the role of epithelial-mesenchymal heterogeneity in driving the failure of anti-cancer therapies.

Multiple non-genetic mechanisms can contribute towards the emergence of phenotypic heterogeneity. The regulatory circuits that govern the phenotypes of different cells often respond differently to the same external cues leading to a phenotypically heterogeneous population. Phenotypes of cells in a population can change stochastically due to the noisy transcription of genes[52] or due to the random partitioning of the parent cell molecules among the daughter cells during cell division[53,54]. Finally, cell-cell communication can cause cells in a population to acquire distinct phenotypes in a non-cell autonomous manner. Each of these three mechanisms has been implicated in the emergence and maintenance of epithelial-mesenchymal heterogeneity. Mathematical and computational modeling approaches have played a key role in determining how these mechanisms can drive epithelial-mesenchymal heterogeneity in populations of cancer cells. Here, we describe mathematical modeling approaches corresponding to each of the three mechanisms.

## Heterogeneity From Cell-to-Cell Variation in Regulatory Kinetics

Large and complex gene regulatory networks underlie different cellular functions such as stem cell differentiation[55,56] and circadian rhythm[57,58]. The dynamical behavior of such large networks can be understood as being driven by a core regulatory circuit with the remaining genes in the circuit being peripheral to circuit dynamics, acting only to alter the signaling status of the core regulatory circuit[59]. The effects of peripheral genes and exogenous signaling can then be modeled as perturbations to the kinetic parameters governing the dynamics of the core regulatory module. This is the approach underlying the framework known as random circuit perturbation or RACIPE[60]. Here, we describe how to use RACIPE for modeling epithelial-mesenchymal heterogeneity.

While multiple signaling pathways have been implicated in controlling EMT and MET, the activities of many of these pathways converge onto a small set of core regulatory players. This set includes the master regulators such as SNAI1, miR-34, miR-200, and ZEB1[8,61]. The effects of different signals modulating EMT and MET can thus be modeled as perturbations to the kinetics of this smaller core regulatory circuit. These perturbations can vary from cell-to-cell, thus representing the differing internal and external signaling states of tumor cells in a population.

We first describe the RACIPE framework using the simple toggle switch as an example. As shown in fig. 2, the toggle switch consists of two transcription factors, $A$ and $B$, which form a mutual inhibitory feedback loop. The dynamics of this circuit can be described using a pair of ordinary differential equations:

$$\frac{d[A]}{dt} = g\ H^S([B], K_B^A, n_B^A, \lambda_B^A) - k_A[A] \quad (1)$$

$$\frac{d[B]}{dt} = g\ H^S([A], K_A^B, n_A^B, \lambda_A^B) - k_B[B] \quad (2)$$

Here, $[A]$ and $[B]$ are the protein expression levels of genes $A$ and $B$ respectively. $g$ and $g$ are the production rates of $A$ and $B$ when no activator or inhibitor is present. $k$ and $k$ are the inherent degradation rates of the two genes. The regulatory action of gene $B$ on gene $A$ is modeled via the shifted Hill function:

$$H\ ([B], K_B^A, n_B^A, \lambda_B^A) = \lambda_B^A + (1 - \lambda_B^A)H^-([B], K_B^A, n_B^A) \quad (3)$$

$$H^-([B], K_B^A, n_B^A) = \frac{1}{1 + \left(\frac{[B]}{K_B^A}\right)^{n_B^A}} \quad (4)$$

$K^A$ is the threshold concentration of $B$, $n^A$ is the Hill coefficient, and $\lambda^A$ is the maximum fold change in the expression level of $A$ that can be caused by the activity of $B$. If $B$ activates $A$, $\lambda^A > 1$. If $B$ inhibits $A$, $0 \leq \lambda^A < 1$. For the toggle switch, $0 \leq \lambda_B^A, \lambda_A^B < 1$. Thus, there are five types of kinetic parameters in the model. Two of them, $g$ and $k$, are associated with each gene. The remaining three, $K$, $n$, and $\lambda$ are associated with each regulatory link. Thus, for a circuit with *10* genes and *25* regulatory interactions, the total number of parameters will be $(2 \times 10) + (3 \times 25) = 95$.

RACIPE performs randomization on all five types of circuit parameters to obtain an ensemble of kinetic models for a given circuit topology. The randomization procedure is such that most biologically realizable possibilities are represented by one of the models in the ensemble. RACIPE further uses two assumptions to obtain a representative ensemble of models. First, the maximum production rate of each gene is fixed, independent of the number and type of interactions that gene is a target of. For a gene with one activator, the maximum production rate, $G$, will be obtained when the activator is highly expressed. Thus, the basal production rate of the gene must be $g = \frac{G}{\lambda}, \lambda > 1$. For a gene with only one inhibitor, the maximum production rate will be obtained in the absence of inhibitor expression. Thus, $G = g$ where $g$ is the basal production rate. This approach can easily be generalized to the case when a gene has multiple activators and inhibitors[60]. RACIPE randomizes the maximum production rate ($G$) and then calculates $g$ using the above-mentioned approach.

The second assumption is that in order for the ensemble of models to be representative of most biological possibilities, each regulatory link in the circuit must have an almost equal chance of being functional and being non-functional. To ensure this, RACIPE chooses the threshold parameters $K$ in such a manner that the steady state concentration of the corresponding regulator in different models within the ensemble is roughly equally likely to be above the threshold parameter (in which case the interaction is functional) and below the threshold parameter (in which case the interaction is non-functional). For a detailed description of how this is achieved, see Huang *et al*.[60]

In the ensemble generated by RACIPE, all models have the same topology but differ in the values of kinetic parameters governing the model dynamics. The dynamics of each model is then numerically simulated multiple times, each time starting with a different set of initial concentrations of the molecules in the circuit. This allows RACIPE to obtain a set of steady states that a given model can generate. Once this has been done for each model in the ensemble, RACIPE obtains a collection of steady states that the given circuit topology can exhibit. Each model in the ensemble generated by RACIPE may be interpreted as representing a single cell. Thus, the collection of steady states obtained by RACIPE will represent an *in silico* gene expression profile obtained for a population of cells. One aspect that should be kept in mind is that a model that can exhibit more than one steady states will be counted more often in the collection of steady states generated by RACIPE as compared to a model that can exhibit only one steady state. Nevertheless, this steady state expression data can be analyzed using familiar methodologies including principal component analysis and hierarchical clustering to gain insight into the different classes of steady states that may be exhibited by a given network topology.

The C language computer code implementing the RACIPE framework is available online on GitHub (https://github.com/simonhb1990/RACIPE-1.0). Once the code has been downloaded, change to the folder or directory where the code files are present and use the *make* command to compile the code files for your system. This will generate a single executable named "RACIPE". This executable takes as input a topology file, extension .topo, which describes the topology of the circuit being analyzed. This must be a plain text file with three tab-separated columns. The first column ("Source") contains the name of the regulator gene. The second column ("Target") contains the name of the gene being regulated. The third and final column ("Type") describe the interaction type, *1* if the interaction is activating and *2* if the interaction in inhibiting. A sample topology file (TS.topo) is available online with the code. Once the topology file for the circuit of interest has been generated, the RACIPE code can be run as follows:

$ ./RACIPE network.topo

Additional input options that may be provided to the code are described in the "README.md" file available with the code. Upon execution, the code generates multiple files. Most important among these are:

1. **Parameter ranges file (.prs extension)** This file contains the ranges of different kinetic parameters.
2. **Parameters file (_parameter.dat extension)** This file contains the kinetic parameters for each model in the ensemble along with the number of steady states obtained for that model.
3. **Solutions files** These files contains the gene expression levels in each of the steady states obtained for different models. Steady state expression levels for models exhibiting different numbers of steady states are stored in different files. For models with only steady state, the file extension is "_solution_*1*.dat".

For models with three steady states, the file extension is "_solution_3.dat". All gene expression values reported in these file are log2 normalized.

Descriptions of other output files can be obtained from the "README.md" file available online with the code.

To determine if epithelial-mesenchymal heterogeneity can emerge from cell-to-cell variation in kinetic parameters as simulated using the RACIPE framework, we used a 26-node circuit (fig. 3; top panel) which was constructed using Ingenuity Pathway Analysis (IPA; QIAGEN Inc.) and literature search[62]. The circuit consists of 17 protein-coding genes and 9 micro-RNAs. The set of protein-coding genes includes transcription factors such as SNAI1, ZEB1, and TWIST1 whose role as master regulators of EMT is well characterized[8]. The set also includes EMT-associated biomarkers such as CDH1 and VIM along with "phenotypic stability factors"[34] such as GRHL2, OVOL2, and ΔNP63α. The collection of steady states that can be exhibited by models with the topology of this EMT circuit was obtained using RACIPE and analyzed using hierarchical clustering (fig. 3; bottom panel). As mentioned previously, this collection of steady states is representative of the gene expression profile of cells in a tumor. The steady states can be broadly classified into four groups on the basis of expression levels of the 26 proteins and micro-RNAs in the EMT circuit. Group 1 exhibits high levels of expression of epithelial phenotype-associated genes including CDH1 along with high levels expression of EMT inhibitors such as GRHL2 and miR-200. This group thus represents cells that exhibit an epithelial phenotype. In group 4, EMT drivers such as SNAI1 and ZEB1 are highly expressed along with high expression of the mesenchymal marker VIM. This group represents cells that exhibit a mesenchymal phenotype. Groups 2 and 3 consist of steady states with co-expression of both epithelial and mesenchymal-associated factors. The expression of epithelial factors in these groups is lower than the expression of these factors in the epithelial group (group 1) and the expression of mesenchymal factors is lower than that in the mesenchymal group (group 4). Groups 2 and 3 thus co-express both epithelial and mesenchymal factors at intermediate levels.

Thus, analysis of a 26-node EMT circuit using the RACIPE framework demonstrates one mechanism by which epithelial-mesenchymal heterogeneity can emerge in a population of cancer cells. Due to the cell-to-cell variation of kinetic parameters driving EMT dynamics, cells can exhibit distinct gene expression profiles that can broadly be grouped into epithelial, mesenchymal, and hybrid E / M classes. The cell-to-cell variation in kinetic parameters is indicative of the differing exogenous signaling states in different cells. While cells in the population exhibit different gene expression profiles, the population does not consist of clones and subclones with cells in each clonal population exhibiting a specific EMT kinetic, i.e. the gene expression profile of a cell is not hereditary and can change in response to changes in the exogenous signaling environment. Note that while our analysis reveals 2 groups of steady states with co-expression of epithelial and mesenchymal factors suggesting that 2 such hybrid E / M states exist, a different analysis technique may reveal a greater number distinct types of hybrid E / M phenotypes. Cells can likely be classified into even greater number of phenotypic groups by incorporating other EMT-associated factors into the circuit topolgy[10,11] which would provide greater resolution as has been reported recently[16]. Finally, the RACIPE framework can easily be used to probe the contribution of each protein and micro-RNA and of each regulatory relationship in driving epithelial-mesenchymal heterogeneity. One can edit the circuit topology file (extension .topo) to add and / or delete EMT-associated factors and regulatory relationships and analyze the expression levels in the collection of steady states obtained for the altered circuit.

## Heterogeneity from Random Partitioning of Molecules During Cell Division

Another scenario in which phenotypic heterogeneity can emerge in a population occurs if cells undergo stochastic changes in their phenotypes. In general, for such stochastic changes to happen, there must exist a mechanism to generate noise and a mechanism to stabilize the decision reached in response to the noise[63]. One mechanism which can generate noise is the random partitioning of molecules (RNAs, proteins, etc.) in the parent cell among the daughter cells at the time of cell division[53,54,64]. This mechanism is likely to be a prominent source of noise in tumors wherein cells divide fast and uncontrollably. While phenotypic fluctuations in cells in response to noise are usually small and transient, the fluctuations can be amplified if the underlying response mechanism exhibits multi-stability, i.e., co-existence of multiple steady states. As described previously[9,13], circuits which drive EMT and MET exhibit multi-stable behavior. Thus, random partitioning of

EMT-associated factors during cancer cell division is likely to be a key contributor towards the emergence of epithelial-mesenchymal heterogeneity.

The schematic representation of a computational model that can be used to probe the role of this mechanism in the emergence of epithelial-mesenchymal heterogeneity is shown in fig. 4. The model[65] builds upon the dynamics of the core regulatory circuit involving SNAIL, ZEB, miR-34a, and miR-200. These transcription factors and micro-RNAs together form a circuit that acts as a ternary switch, responding to the signaling pathways driving EMT and MET[9]. Stable steady states of this circuit can be mapped to different EMT-associated phenotypes— epithelial, mesenchymal, and hybrid E / M— on the basis of expression levels of ZEB (fig. 4). To see the effect of random portioning on the phenotypic composition of the population, we here consider a population of cancer cells with each cell carrying a copy of this EMT regulatory circuit. Since this regulatory circuit does not involve cell-cell communication, the dynamics of the regulatory circuit within each cell in the population can be simulated independent of other cells in the population. The dynamics of EMT regulation at the single-cell level are simulated using ordinary differential equations which have been described previously[9]. At the population level, there are two types of events that can take place. One is cell death during which a cell is simply removed from the population. The other is cell division.

When a cell divides, the molecules present in the parent cells are randomly partitioned among the daughter cells[53,54,64]. Thus, each daughter cell receives a copy of the EMT regulatory circuit. However, due to the random partitioning of molecules, the concentrations of a molecular species in the two daughter cells can be different from each other and different from the concentration of that species in the parent cell. Let $I$ represent the multiple signaling pathways that converge onto the core EMT regulatory circuit. We here consider noise in the partitioning of $I$ as the dominant perturbation to EMT regulation in the daughter cells. The concentrations of $I$ in the daughter cells are given as:

$$I^{daughter} = I^{parent}_{sig} + \eta N(0,1) \qquad (4)$$

Here, $N(0,1)$ is a standard normal distribution and $\eta$ is a model parameter which determines the variance of the noise distribution. Due to the perturbation in the concentration of $I$, a daughter cell may acquire a phenotype different from that of the parent cell. The population can then become phenotypically heterogeneous over time.

Since the dynamics of EMT regulation is much faster as compared to the time scale at which cell division and cell death events take place, the model dynamics can be simulated in a multi-scale manner. Population-level dynamic, i.e. cell division and cell death, are simulated in a stochastic manner using Gillespie's algorithm[66]. Between each cell division and cell death event, the concentrations of RNAs and transcription factors within each cell are updated using ordinary differential equations. Previous studies have shown that different EMT-associated phenotypes can exhibit different rates of cell division [67–69]. However, one may consider a simpler case with equal division and death rates for all three cell types. In addition, to incorporate the effect of limited availability of nutrients in the tumor microenvironment, a logistic model of growth with a fixed carrying capacity can be used.

Dynamics of the model can be simulated as follows:

1. Choose an initial population size and randomly assign concentrations of molecules in the EMT regulatory circuit to different cells in the population. The concentrations are drawn from log-normal distributions such that the median concentration of each molecular species is within the range for which the regulatory circuit exhibits multi-stable dynamics.
2. Using Gillespie's algorithm[66], update the number of cells in the population. In case of a cell death event, that cell is removed from the simulation and thus the population. In case of a cell division event, $I$ concentrations in the daughter cells are updated using equation *4*.
3. At the end of the Gillespie update, the concentrations of molecules in each cell in the population is updated. Let $\Delta t$ be the time interval between the last Gillespie update and the current one. Then, the concentrations of molecules can be updated by integrating the ordinary differential equations for the EMT regulatory circuit[9] over the time period $\Delta t$.

Computer code for simulating the model dynamics can be downloaded from GitHub (https://github.com/st35/cancer-EMT-heterogeneity-noise).

We simulated the model dynamics for populations with different initial phenotypic compositions. Fig. 5 shows how epithelial-mesenchymal heterogeneity can emerge in a phenotypically homogeneous population over a period of two weeks. While epithelial and mesenchymal populations exhibit fairly stable phenotypic compositions, a hybrid E / M population can quickly give rise to a mixed population with both epithelial and mesenchymal cells. Such behavior has been confirmed in populations of mouse prostate cancer cells[17] and comparison of experimental dynamics with the predictions from the model is shown in fig. 5 (bottom panel).

The model thus shows that random partitioning of parent cell proteins and RNAs among the daughter cells can generate epithelial-mesenchymal heterogeneity in a population of cancer cells. Arising from cell division, this heterogeneity can arise and be propagated from a small population, such as the one left after an anti-cancer regime. Note that the model proposed here is not sensitive to the choice of the core EMT / MET regulatory circuit. Any circuit topology can be used within the framework of this model as long as the circuit dynamics is multi-stable which is a key feature of EMT regulation.

## Heterogeneity from Cell-Cell Communication via Notch Signaling

In addition to the regulatory mechanism at the single-cell level, cell-cell communication also plays a major role in modulating EMT[6,7]. Notch signaling[70,71] is one such mechanism which operates via the binding of Notch, a transmembrane receptor, to a ligand expressed on the surface of a neighboring cell. This binding event triggers the cleavage of the Notch intracellular domain (NICD). NICD is then released into the cytoplasm where it can act as a transcriptional cofactor thereby promoting or inhibiting the expression of certain genes[70]. Notch signaling between neighboring cells can create varied spatial patterns in a population. The pattern type depends on the type of Notch ligands that are active in the population. NICD activates the expression of Delta ligands and promotes the expression of ligands of the Jagged family. Notch-Delta signaling leads to neighboring cells acquiring distinct phenotypes— the cell expressing high levels of the Notch receptor and low levels of Delta ligands acts as the "sender" cell while the neighboring cell with low levels of Notch expression and high expression levels of Delta ligands acts as the "receiver" cell[72] ("lateral inhibition"; fig. 6 (top panel)). Notch-Jagged signaling, on the other hand, leads to neighboring cells acquiring the same phenotype which is characterized by the co-expression of Notch receptors and Jagged ligands[73] ("lateral induction"; fig. 6 (bottom panel)).

The role of Notch signaling in EMT regulation arises from the coupling between the Notch signaling machinery and the core regulatory circuit that drives EMT (fig. 7; top panel). miR-34 can post-transcriptionally inhibit the expression of Notch receptors and that of Delta ligands. miR-200 similarly inhibits the expression of Jagged ligands. Further, NICD promotes the expression of SNAIL, thereby acting as an EMT promoter[4]. Due to the cross-talk between the Notch signaling and EMT circuits, the spatial patterns that emerge from Notch signaling translate into spatial patterning in the expression of epithelial and mesenchymal markers in a population of cells. In general, since NICD is an EMT promoter and Notch-Delta signaling leads to neighboring cells acquiring distinct phenotypes, Notch-Delta signaling leads to a spatial expression profile wherein hybrid E / M and mesenchymal cells are surrounded by epithelial cells. On the other hand, Notch-Jagged signaling can lead to the emergence of spatial clusters of hybrid E / M and mesenchymal cells due to the tendency of neighboring cells to acquire the same phenotype in the presence of Notch-Jagged signaling.

The spatial expression of epithelial and mesenchymal factors in a population in the presence of Notch signaling can be probed using ordinary differential equations to model the behavior of coupled Notch signaling and EMT circuits. The methodology differs from previous models of EMT regulation in that the dynamics of the circuit within each cell depends not only the concentrations of molecules within the cell but also on the concentrations of molecules, particularly Notch receptors and ligands, on neighboring cells that are in direct contact with the given cell. Therefore, before simulating Notch signaling mediated dynamics, one must choose a suitable spatial lattice wherein each lattice position is occupied by a single cell. This is essential in order to properly identify the neighboring cells for each cell in the population. We will not describe the mathematical form of the ordinary differential equations here since these equations have been described in detail previously[4].

Fig. 7 (bottom panel) shows the spatial patterns that emerge via Notch signaling between cells occupying a hexagonal lattice wherein each cell communicates with six neighboring cells in the population. The results indicate that spatial epithelial-mesenchymal heterogeneity can emerge in a population of cancer cells due to the activity of the Notch signaling mechanism. Cells with differing expression levels of epithelial and mesenchymal factors can be spatially organized in distinct patterns in different contexts. While Notch-Delta signaling leads to a "salt-and-pepper" patterning wherein hybrid E / M and mesenchymal cells are surrounded by epithelial cells, Notch-Jagged signaling leads to the emergence of clusters of these cell types. The spatial organization of epithelial-mesenchymal heterogeneity is a distinguishing feature of this mechanism for emergence of heterogeneity. Neither cell-to-cell variation in kinetic parameters governing EMT regulation nor random partitioning of molecules during cell division can lead to such behavior. Spatial heterogeneity in the abundance of different phenotypes is a characteristic of tumors[74]. For example, mesenchymal cancer stem cells are abundant near the tumor-stroma boundary while cancer stem cells exhibiting a hybrid E / M phenotype tend to localize in the interior of the tumor[75]. The cell-cell communication-dependent mechanism for the generation of phenotypic heterogeneity described here can be used to understand and describe such features of the tumor microenvironment[76].

## Modeling the Coupling Between EMT and Stemness in Cancer Cells

Across cancer types, subpopulations of tumor cells that exhibit stem-cell like behavior, i.e., an increased capacity to repopulate tumors, have been observed[77]. These cancer stem cells (CSCs), often inherently resistant to anti-cancer therapies, can not only repopulate the tumor post-therapy but also re-create the intra-tumoral heterogeneity exhibited by the original tumor. The connection between epithelial-mesenchymal transition and the appearance of stem cell-like properties in cancer cells has been studied for a long time. Initial studies argued that tumor cells must undergo a complete EMT in order to exhibit traits of CSCs[78,79]. This proposition was consistent with the then prevalent perception of EMT as a binary process. Later studies showed that EMT / MET and cancer cell stemness are both highly dynamic processes. Cancer cells can exhibit hybrid E / M phenotypes and inter-convert between the different EMT-associated phenotypes. Similarly, cancer cells can switch between CSC and non-CSC phenotypic states, maintaining a dynamic equilibrium in a population of cancer cells[80–83]. Due to these developments, a more nuanced picture of the EMT-stemness connection has emerged wherein all EMT-associated phenotypes— epithelial, mesenchymal, and hybrid E / M— can exhibit stemness properties depending on the strength of coupling between the modules regulating EMT and stemness.

In cancer cells, stemness is regulated by a 2-component decision making circuit (fig. 8) wherein LIN28 and let-7, a micro-RNA, form a mutual inhibitory loop. NF-$\kappa$B activates the expression of both LIN28 and let-7 and thus acts as an input to this regulatory module. Both LIN28 and let-7 can also activate their own expression. The dynamics of this regulatory circuit can be modeled using ordinary differential equations (ODEs) as has been done previously for the EMT regulatory circuit. These ODEs have been described in detail elsewhere[19]. The ODEs can be integrated numerically to obtain the steady state expression levels of LIN28 and let-7 for different concentrations of NF-$\kappa$B. Three distinct phenotypes are evident from this analysis— high LIN28, low LIN28, and intermediate LIN28. LIN28 activates the expression of the pluripotency marker OCT4[84] and the stem cell state is characterized by the expression of OCT4 within a range— both very low and very high levels of OCT4 expression lead to the loss of stemness[85–88]. Thus, only cells with such intermediate levels of OCT4 expression can acquire a cancer stem cell phenotype.

The stemness regulatory module couples with the EMT regulatory module via two micro-RNA mediated regulatory interactions. miR-200 inhibits the expression of LIN28 post-transcriptionally. Similarly, let-7 inhibits the expression of the EMT-driver ZEB (fig. 8). These interactions can easily be included in the ODE-based models of EMT and stemness regulation to couple the two regulatory units[89]. Since both very low and very high levels of OCT4 expression lead to loss of stemness, to determine if a certain EMT-associated phenotype can acquire stemness, one can define a "stemness window"— range of expression levels of OCT4 for which a cell can acquire stemness. EMT-associated phenotypes that overlap with this stemness window can then acquire stemness. This overlap, and thus the set of EMT-associated phenotypes that can acquire stemness can be modulated by varying the strength of coupling between the two regulatory units. This coupling is modeled via two parameters— $\alpha_1$, the maximum fold change in the expression level of LIN28 that miR-200 can cause, and

$\alpha_2$, the maximum fold change in the expression level of ZEB that let-7 can cause. Since miR-200 inhibits LIN28 and let-7 inhibits ZEB, $0 \leq \alpha_1, \alpha_2 \leq 1$. A fold change close to *1* indicates weak coupling while a fold change close to *0* indicates strong coupling.

When there is no coupling between the two regulatory units ($\alpha_1 = 1$, $\alpha_2 = 1$), cells can exhibit three distinct phenotypes associated with the EMT circuit (epithelial, mesenchymal, and hybrid E / M) provided the concentration of the EMT-driver SNAIL is within the range for which the EMT circuit can exhibit tri-stability. Cells can further exhibit three distinct phenotypes corresponding to the stemness circuit (low LIN28, high LIN28, and intermediate LIN28). Thus, *9* (*3 × 3*) total phenotypes are possible. This number decreases when the strength of the coupling between the circuits is increased. Which of the EMT-associated phenotypes exist within the stemness window depends on the relative values of $\alpha_1$ and $\alpha_2$. When both $\alpha_1$ and $\alpha_2$ are close to *1* (weak coupling), all three EMT-associated phenotypes lie within the "stemness window" and thus can acquire stemness. Upon decreasing $\alpha_1$, the stemness window shifts towards the mesenchymal phenotype. Epithelial cells can no longer acquire stemness in this scenario. When $\alpha_2$ is decreased while keeping $\alpha_1$ close to *1*, the stemness window shifts towards the epithelial phenotype and mesenchymal cells cannot acquire stemness with such a coupling between the regulatory units. The different scenarios have been illustrated in fig. 9.

The total number of phenotypes that may be exhibited by cells in a population will further depend on the concentration of SNAIL. For example, when the SNAIL concentration is very high, cells can only exhibit the mesenchymal phenotype. These mesenchymal cells can then acquire stemness provided the "stemness window" overlaps with the mesenchymal phenotype. Similarly, very low concentrations of SNAIL will lead to cells in the population exhibiting only the epithelial phenotype. In such a scenario, two distinct phenotypes may be acquired by tumor cells in the population— epithelial stem-like and epithelial non-stem-like. The number of phenotypes exhibited can further be tuned by varying concentrations of NF-$\kappa$B which activates the expression of both LIN28 and let-7. Very low or very high NF-$\kappa$B concentrations, for example, will cause cells to lose their ability to acquire and maintain stemness due to very low and very high OCT4 expression levels respectively.

The coupling of EMT and stemness regulatory modules thus allows for the existence of a myriad of phenotypes. Tumor cells in a population may exhibit all or some of these phenotypes depending on the signaling profile. Coupled with the spatial heterogeneity of signaling states within a tumoral mass, subpopulations exhibiting different phenotypic profiles can exist in different parts of the tumor. The EMT regulatory circuit in cancer cells is further coupled with other regulatory modules including the Notch signaling module. Such additional couplings can further increase the number of phenotypes that can be exhibited by cells in a population in a manner similar to the EMT-stemness coupling described above. Additionally, since Notch signaling leads to the emergence of spatial patterns in the distribution of different phenotypes, EMT-Notch-stemness coupling can lead to the localization of different stemness associated phenotypes in different parts of the tumor microenvironment[76].

# Conclusion

Here, we have presented EMT from the lens of computational systems biology where the focus is on the emergent properties of the underlying regulatory network, instead of those of individual nodes in the network. We have highlighted various examples of how physics / engineering / mathematics driven approaches can reveal unprecedented insights into various aspects of EMT dynamics, such as multistability, reversibility / irreversibility, symmetry (or not) in EMT / MET, the effects of non-cell autonomous mechanisms in EMT / MET, and finally the connection of EMT / MET with other cellular traits such as stemness. The *in silico* models presented here have their own strengths, limitations, and assumptions, just as is the case with any *in vitro*, *in vivo*, or *ex vivo* model. The examples presented here emphasize how an iterative crosstalk between mathematical modeling and experimental biology can help decode plasticity and heterogeneity in EMT / MET.

# Acknowledgements

This work was supported by the Ramanujan Fellowship awarded to M.K.J. by SERB, DST, Government of India (SB/S2/RJN-049/2018).

Figures

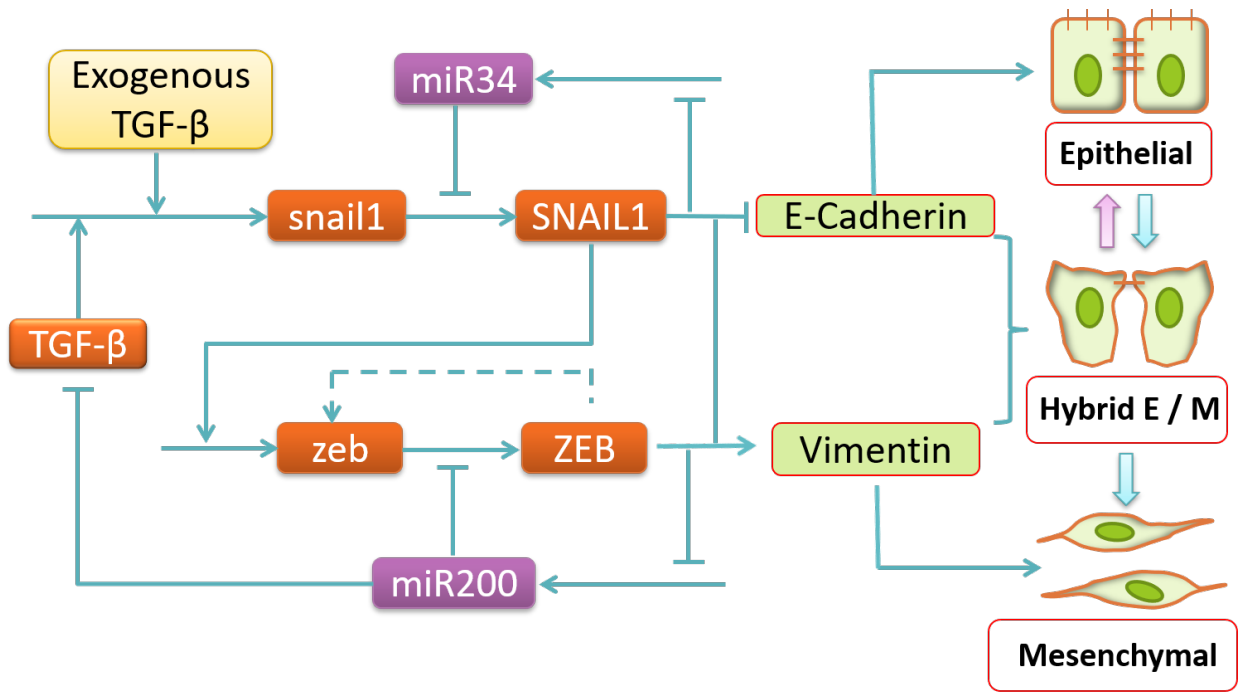

**Figure 1** Core EMT regulatory network that leads to epithelial, hybrid E / M, and mesenchymal phenotypes (adapted from[13]). Point arrows represent activation, blunt-end arrows represent inhibition, and the dashed lines represent links first proposed in the modeling study by Lu *et al.*[9].

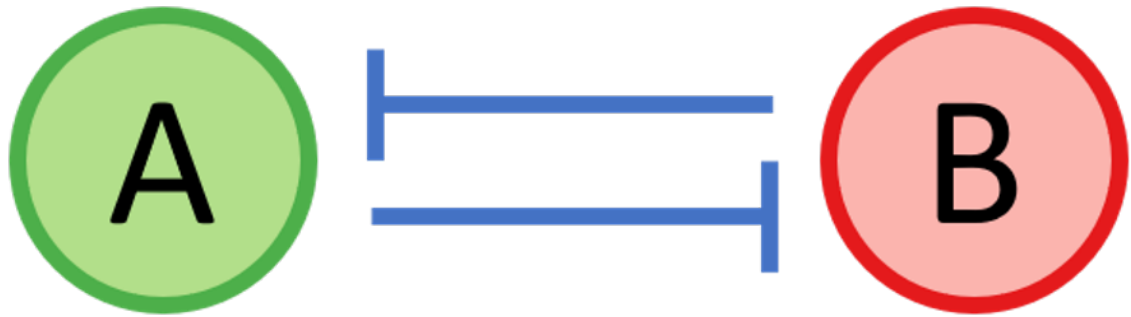

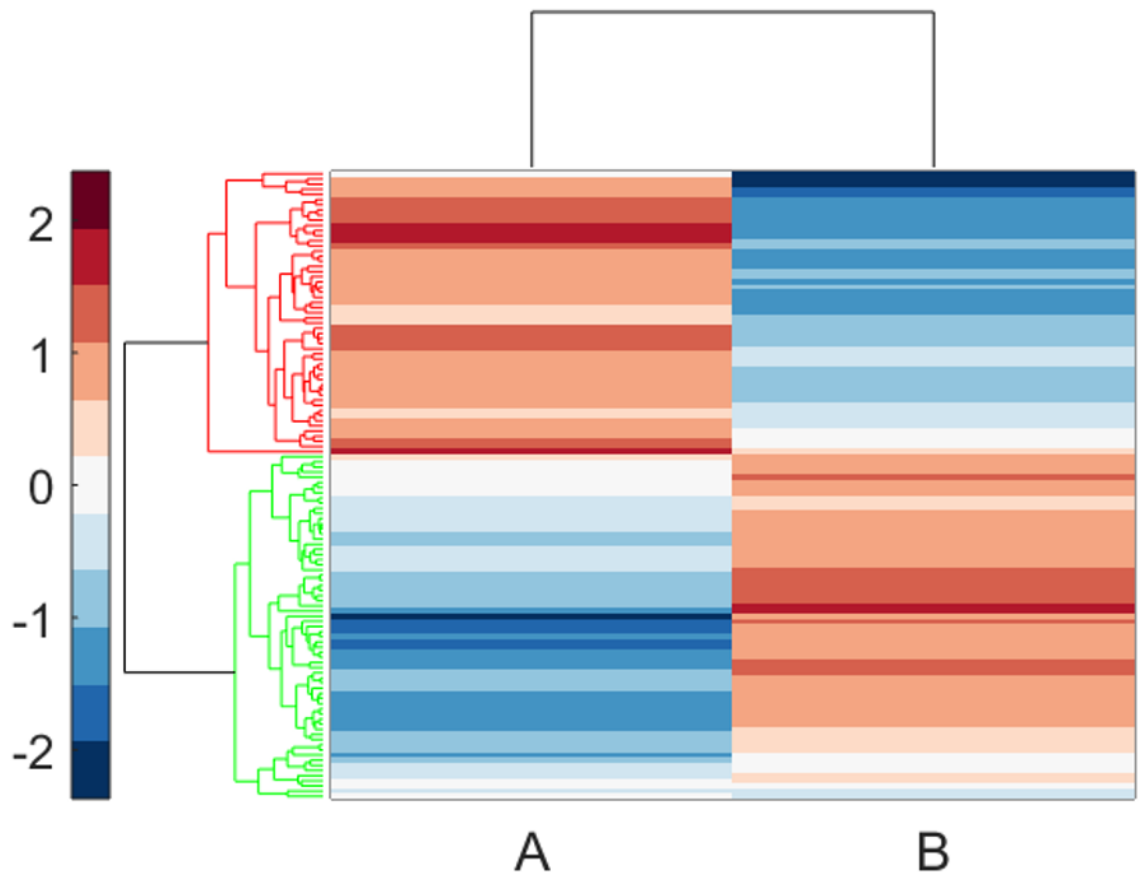

**Figure 2** Transcription factors *A* and *B* with a mutual inhibitory feedback loop (top). RACIPE was used to generate *100* kinetic models corresponding to this topology. A total of *122* distinct steady states were obtained— *78* kinetic models exhibited only one steady state while *22* kinetic models exhibited two steady states. Hierarchical clustering of this collection of steady states (bottom) revealed that these steady states can be divided into two phenotypic classes: high A, low B (highlighted in red) and low A, high B (highlighted in green). Thus, in a population wherein each cell carries a copy of this circuit, cells can exhibit two distinct phenotypic states. Hierarchical clustering was carried out using the Z-scores of the log*2* transformed expression levels.

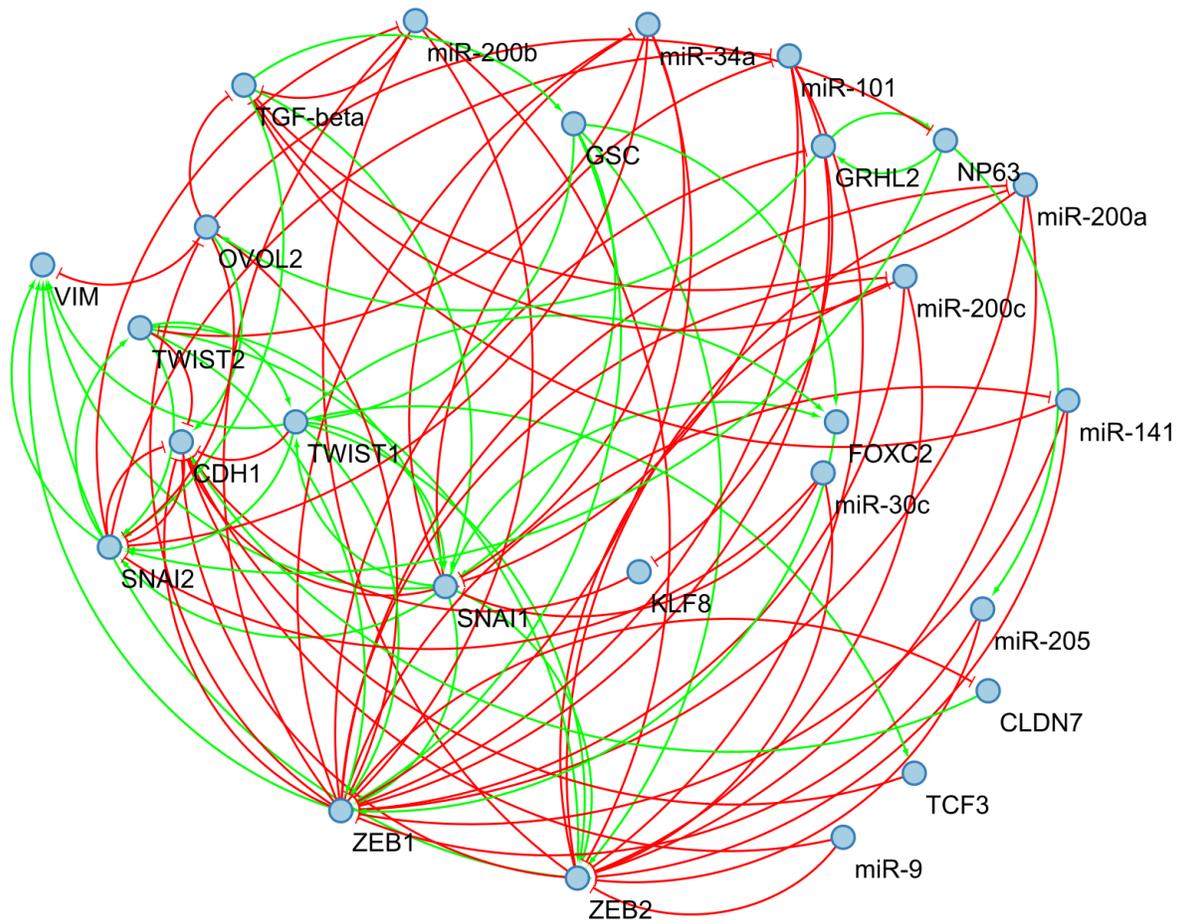

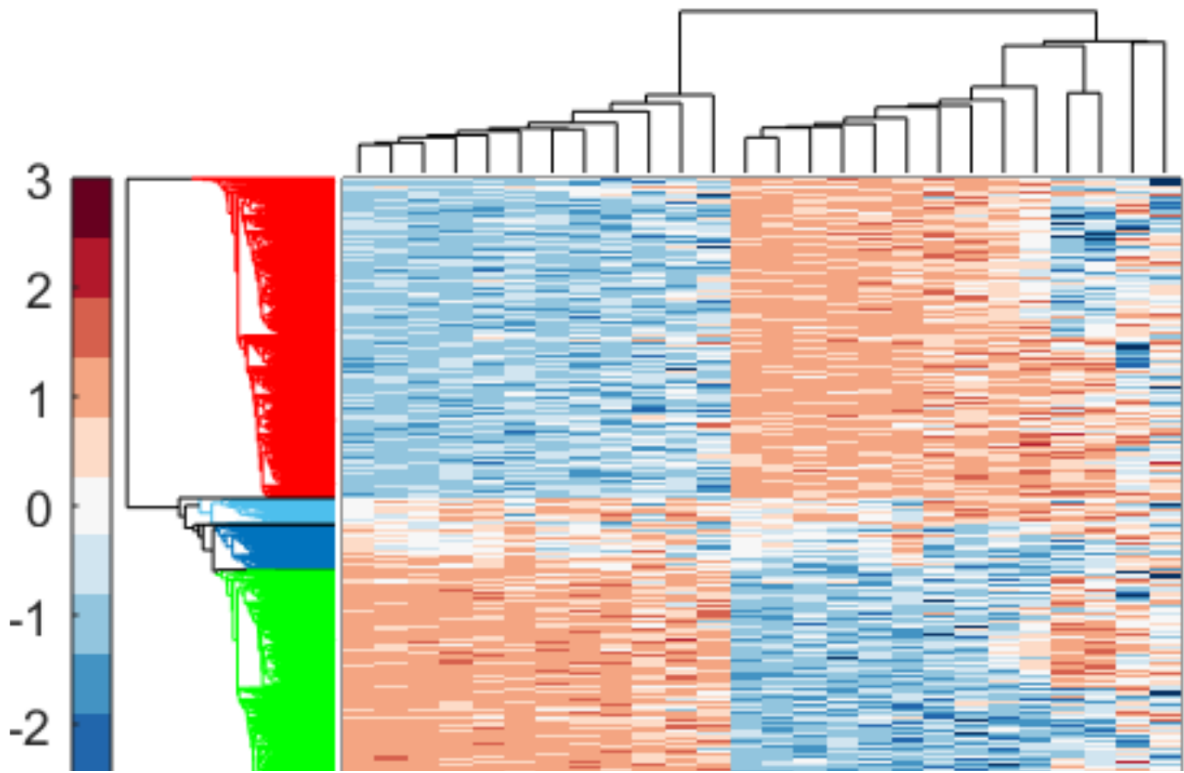

**Figure 3** The *26*-node EMT topology (top). RACIPE was used to generate *5000* kinetic models corresponding to this topology. A total of *13486* steady states were obtained via numerical integration of the ODEs in these kinetic models. Using hierarchical clustering, these steady states were grouped into four phenotypic classes (bottom)— epithelial (red), mesenchymal (green), and two hybrid E / M phenotypic classes (light blue and dark blue). Hierarchical clustering was carried out using the Z-scores of the log2 transformed expression levels.

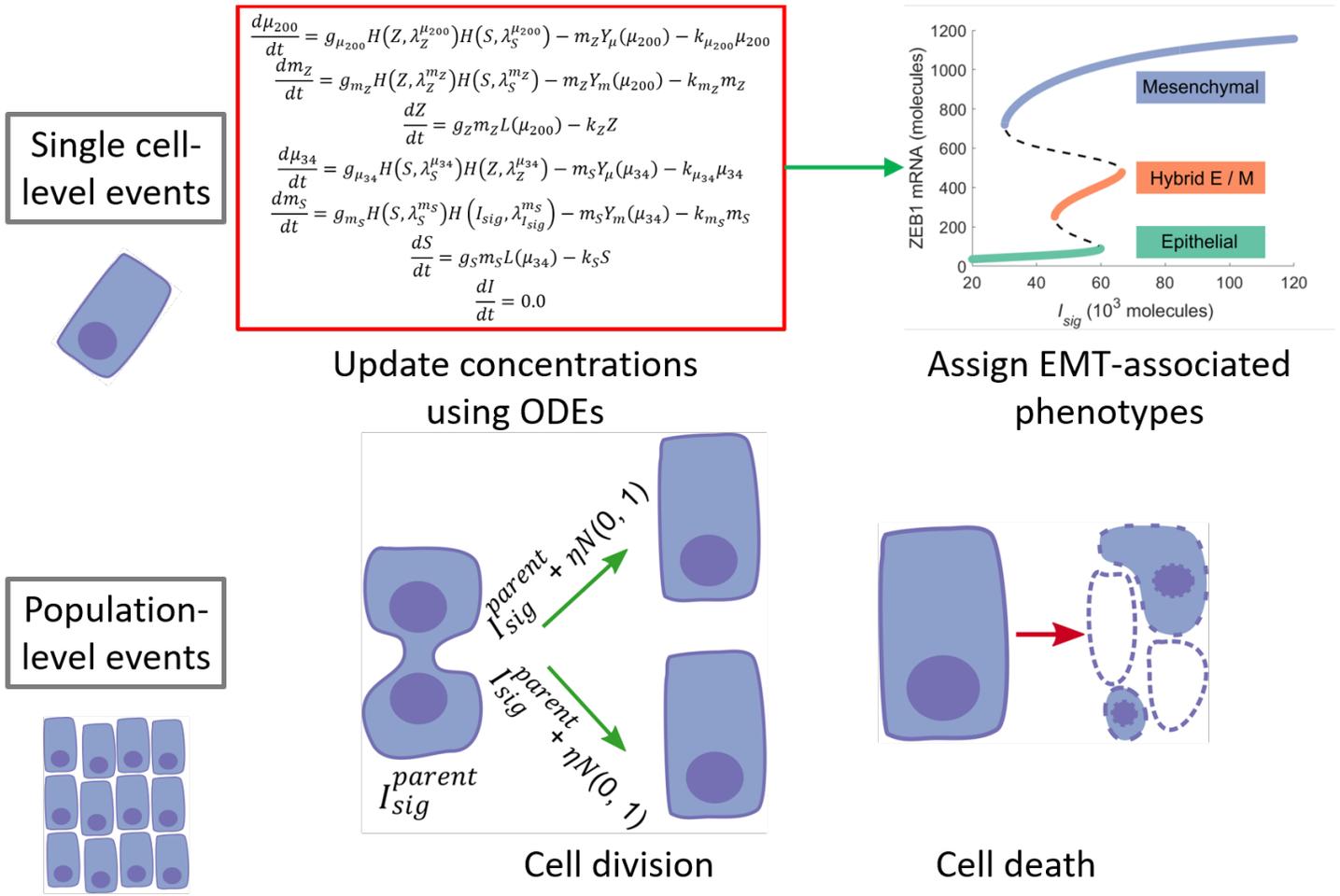

**Figure 4** A schematic representation of the model to investigate how epithelial-mesenchymal heterogeneity can arise from the random partitioning of proteins and RNAs during cell division. Figure adapted from Tripathi *et al*[65].

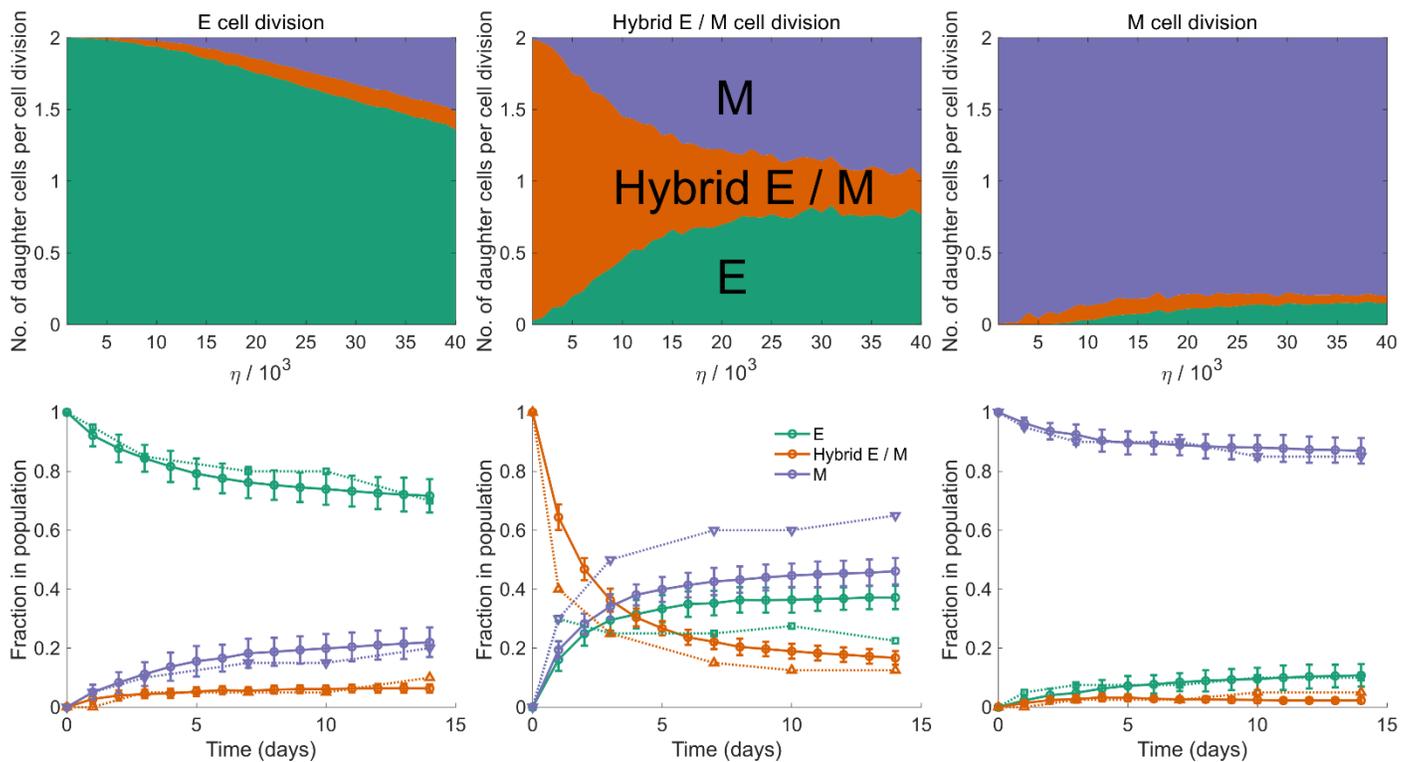

**Figure 5** (Top) Average number of epithelial, mesenchymal, and hybrid E / M daughter cells generated during the division of an epithelial cell (left), a hybrid E / M cell (middle), or a mesenchymal cell (right). Daughter cells can exhibit a phenotype distinct from that of the parent cell due to the random partitioning of $I$ during cell division. (Bottom) Change in the fraction of different phenotypes in a population of cancer cells when starting with a purely epithelial (left), a purely hybrid E / M (middle), or a purely mesenchymal population on day *0*. Solid lines indicate the predictions from the proposed model. Dotted lines indicate the behavior for a population of mouse prostate cancer cells re-plotted from Ruscetti *et al*[17]. Figure adapted from Tripathi *et al*[65].

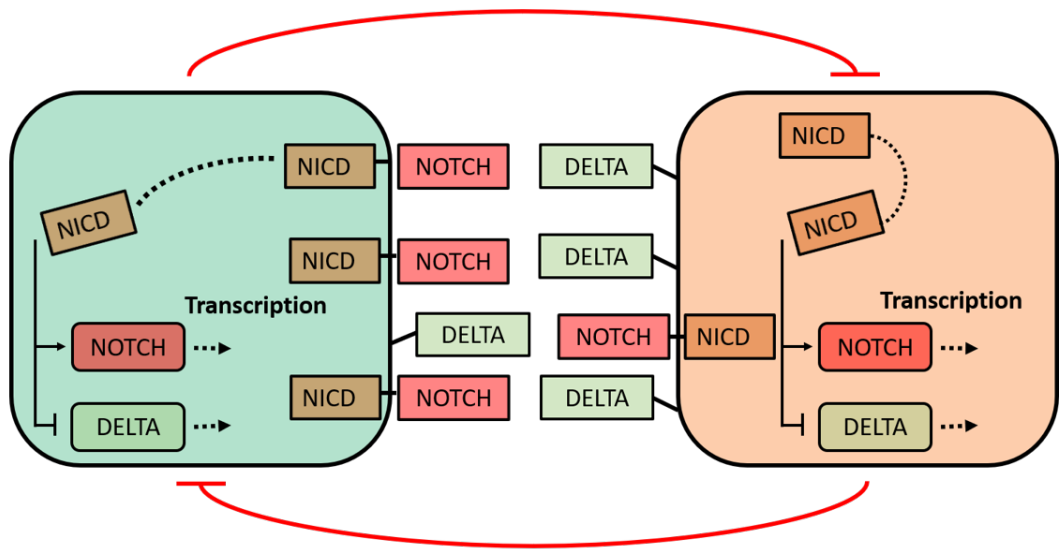
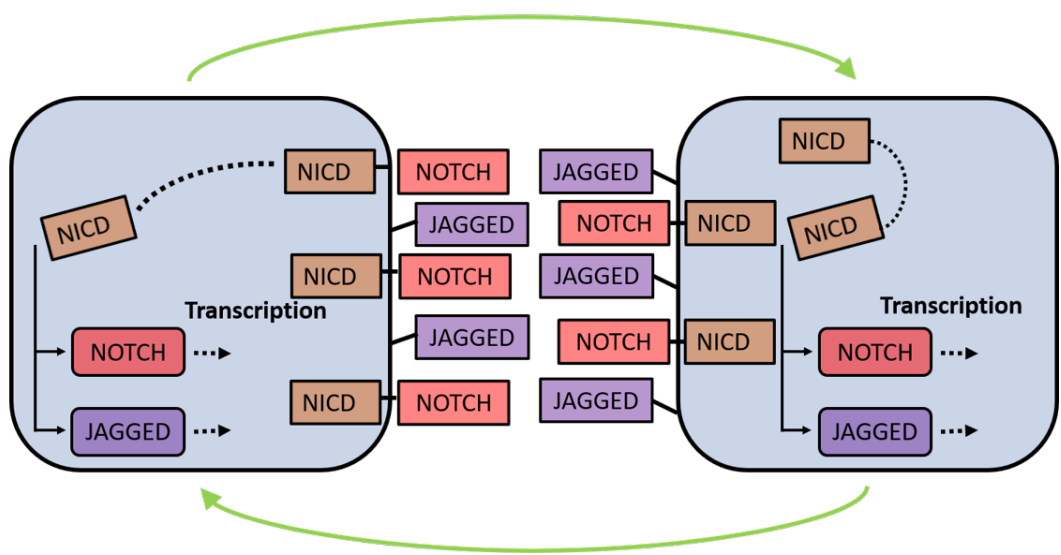

**Figure 6** Two types of Notch signaling-mediated coupling between neighboring cells. In the presence of Notch-Delta signaling (top), neighboring cells form a mutual inhibitory feedback loop causing them to exhibit distinct phenotypes. One of the cells acts as the receiver (green cell in the top panel) with high Notch, low Delta expression. The other cell acts as the sender (orange cell in the top panel) with low Notch, high Delta expression. On the other hand, in the presence of Notch-Jagged signaling (bottom), neighboring cells form a mutual excitatory feedback loop causing them to acquire the same phenotype. Each cell acts both as a sender and a receiver and both cells co-express Notch receptors and Jagged ligands.

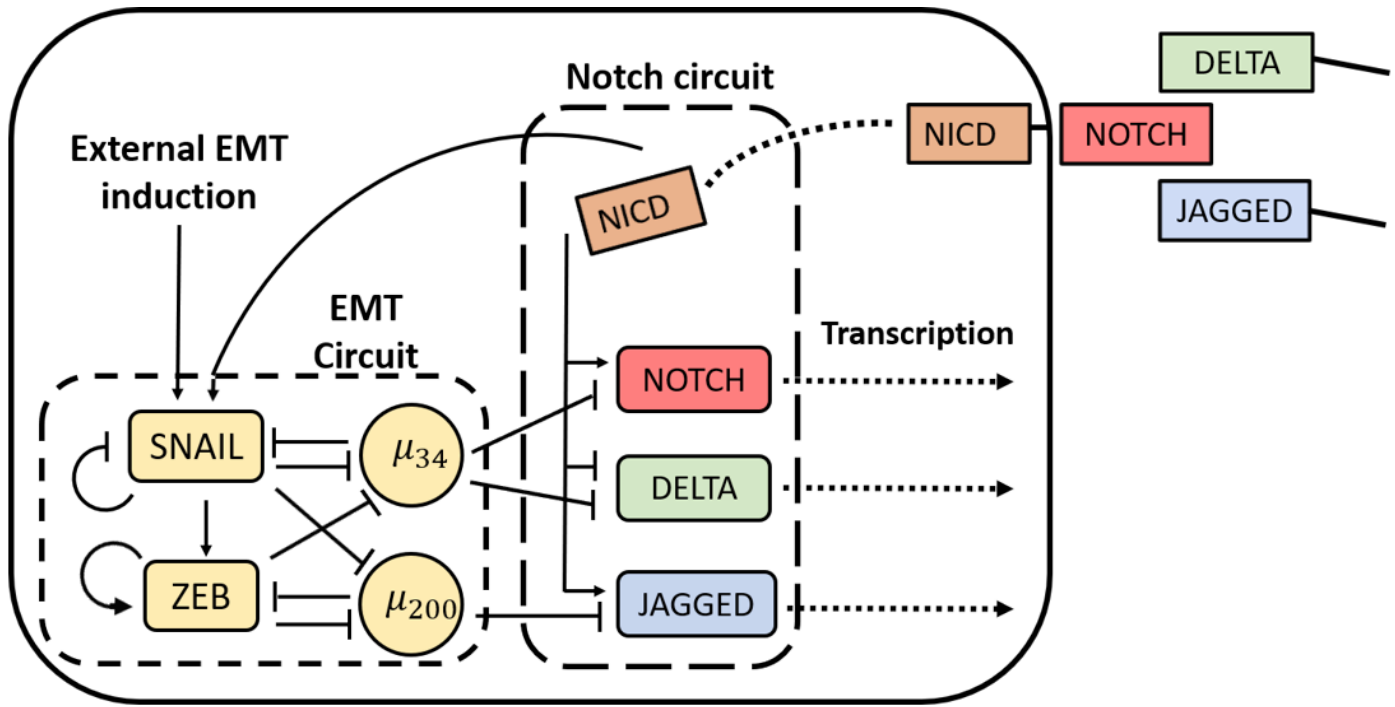
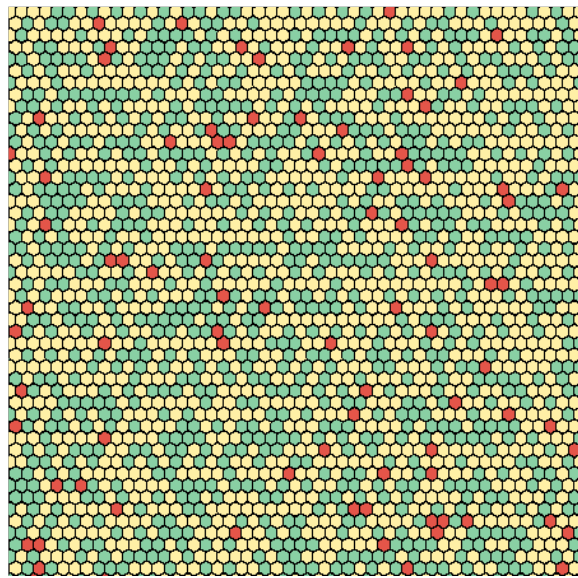
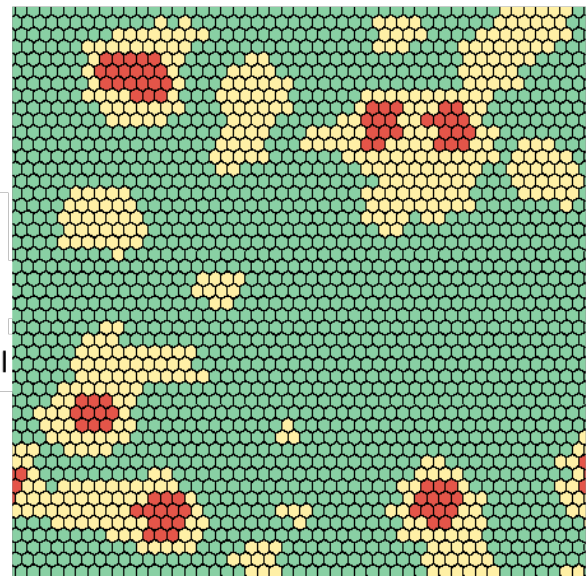

**Figure 7** (Top) Coupling between Notch-Delta-Jagged signaling and EMT regulation. (Bottom) Spatial heterogeneity in the expression of epithelial and mesenchymal markers in the presence of Notch-Delta signaling (left) and in the presence of Notch-Jagged signaling (right).

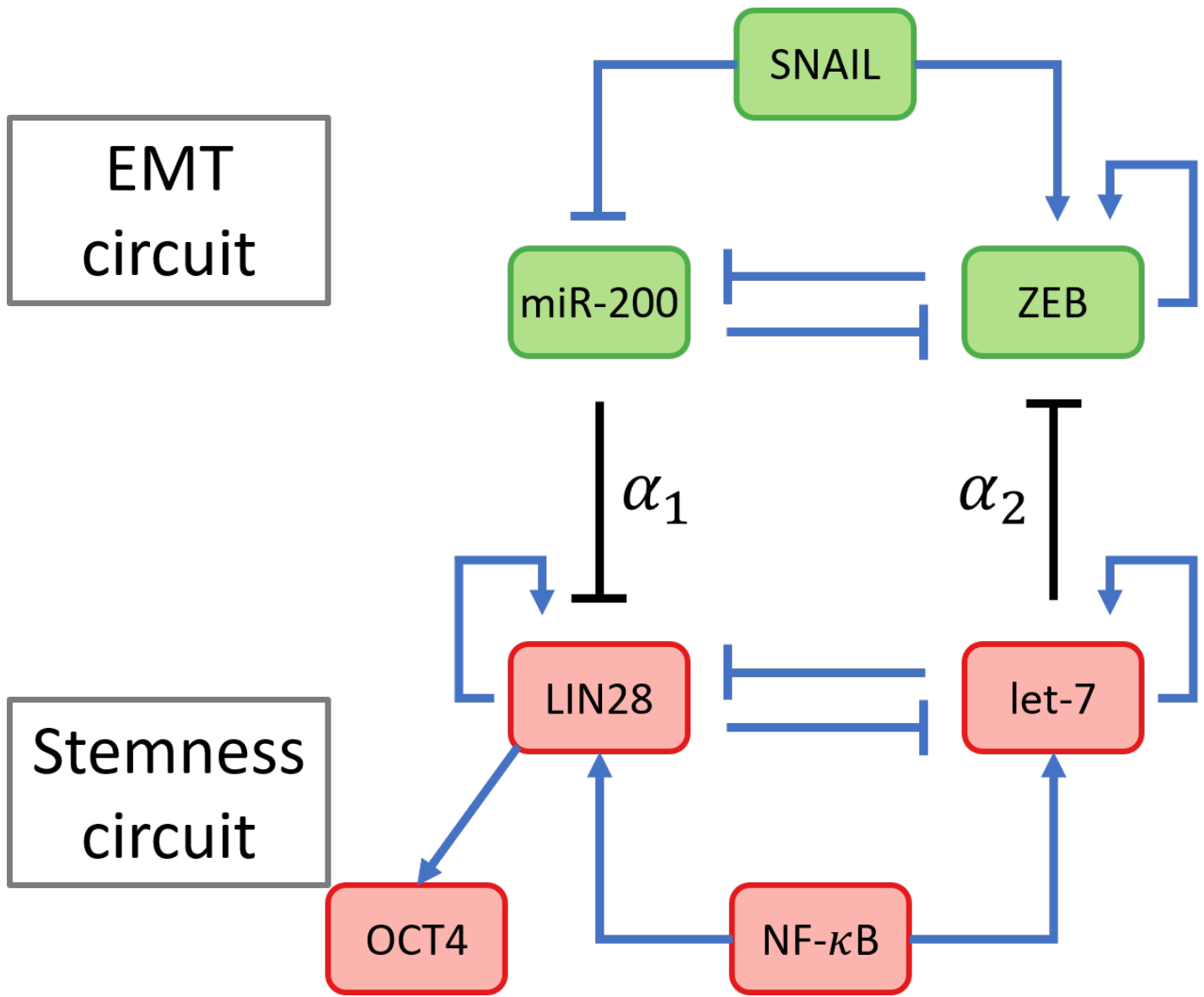

**Figure 8** Coupling between the circuits regulating EMT and stemness. The strength of coupling between the two circuits is governed by the parameters $\alpha_1$ and $\alpha_2$. $\alpha_1$ is the maximum fold change in the rate of production of LIN28 that miR-200 can cause while $\alpha_2$ is the maximum fold change in the rate of ZEB production that let-7 can cause. Since both coupling interactions are inhibitory, $0 \leq \alpha_1, \alpha_2 \leq 1$ with $\alpha_1, \alpha_2 \sim 1$ indicating weak coupling.

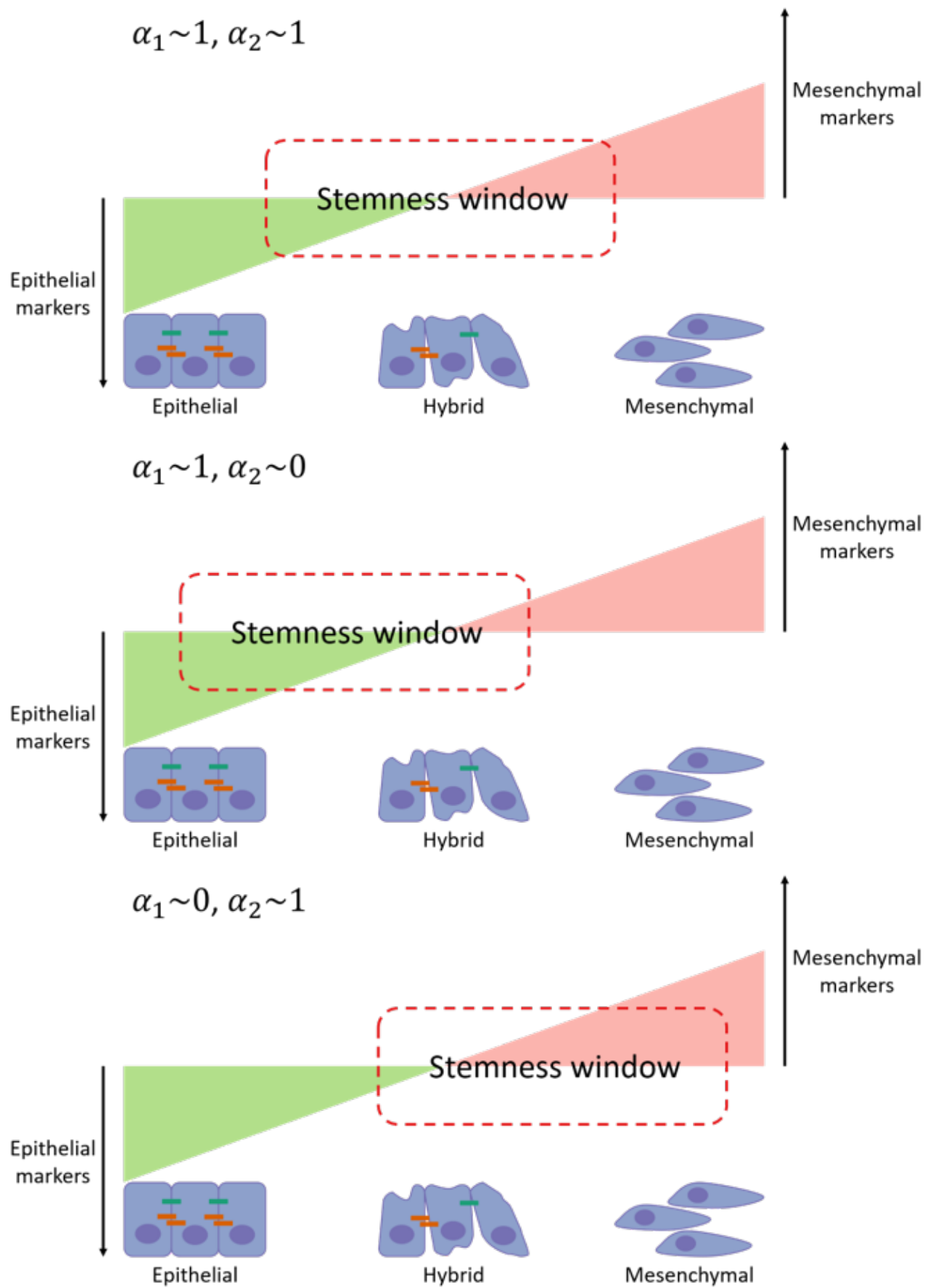

**Figure 9** The coupling parameters determine the overlap of the stemness window (expression of OCT4 within a range) with the spectrum of EMT-associated phenotypes. The overlap determines which of the phenotypes can acquire stemness. In the top panel, all three phenotypes can acquire stemness. In the middle panel, only epithelial and hybrid E / M phenotypes can acquire stemness. In the bottom panel, only hybrid E / M and mesenchymal phenotypes can acquire stemness.